\documentstyle[11pt,aaspp4] {article}

\begin{document}
 
\title{'Round the Clock Observations of the Q0957+561 A,B Gravitationally
Lensed Quasar}
 
\author{
Wesley N. Colley\altaffilmark{1},
Rudolph E. Schild\altaffilmark{1},
Cristina Abajas\altaffilmark{2},
David Alcalde\altaffilmark{2},
Zeki Aslan\altaffilmark{3,4},
Rafael Barrena\altaffilmark{2,5},
Vladimir Dudinov\altaffilmark{6},
Irek Khamitov\altaffilmark{4},
Kjetil Kjernsmo\altaffilmark{7},
Hyun Ju Lee\altaffilmark{8},
Jonghwan Lee\altaffilmark{8,9},
Myung Gyoon Lee\altaffilmark{8},
Javier Licandro\altaffilmark{10},
Dan Maoz\altaffilmark{11,12},
Evencio Mediavilla\altaffilmark{2},
Ver\'onica Motta\altaffilmark{2,13},
Jose Mu\~noz\altaffilmark{2},
Alex Oscoz\altaffilmark{2},
Miquel Serra-Ricart\altaffilmark{2},
Igor Sinelnikov\altaffilmark{6},
Rolf Stabell\altaffilmark{7},
Jan Teuber\altaffilmark{14}, and
Alexander Zheleznyak\altaffilmark{6}}

\altaffiltext{1}{Harvard-Smithsonian Center for Astrophysics, 60 Garden St.,
MS-20, Cambridge, MA 02138}

\altaffiltext{2}{Instituto de Astrofisica de Canarias, Via Lactea, E-38200
La Laguna, Tenerife Canary Islands, Spain}

\altaffiltext{3}{Akdeniz University, Physics Department, 07058 Antalya, Turkey}

\altaffiltext{4}{Tubitak National Observatory, Akdeniz Universitesi Kampusu,
07058 Antalya, Turkey}

\altaffiltext{5}{Dpto. de Astrof\'{\i}isica de la Universidad de La Laguna,
38200 La Laguna, Tenerife, Spain}

\altaffiltext{6}{Kharkov National University, Astronomical Observatory, Sumska
Str. 35, Kharkov, 61022, Ukrania}

\altaffiltext{7}{University of Oslo, Institute of Theoretical Astrophysics, PO
Box 1029, Blindern, Oslo, N-0315, Norway} 

\altaffiltext{8}{Seoul National University, Astronomy Program, SEES, Seoul,
151-742, Rep. of Korea} 

\altaffiltext{9}{DACOM Corporation, Internet Technology Division, DACOM
Bldg., 706-1, Yeoksam-dong, Kangnam-ku, Seoul 135-610, Rep. of Korea}

\altaffiltext{10}{Telescopio Nazionale Galileo, 38700 Santa Cruz de La Palma,
P.O. Box 565, Tenerife, Spain}

\altaffiltext{11}{School of Physics and Astronomy and Wise Observatory, Tel
Aviv University, Tel Aviv 69978, Israel}

\altaffiltext{12}{Dept. of Astronomy, Columbia University, 550 W. 116th St.,
New York, NY 10027} 

\altaffiltext{13}{Departmento de Astronom\'{\i}a, Facultad de Ciencias,
Universidad de la Rep\'ublica, Uruguay}

\altaffiltext{14}{Danish Library for Natural and Medical Sciences, Copenhagen,
Denmark}

 
\begin{abstract}

An observing campaign with 10 participating observatories has undertaken to
monitor the optical brightness of the Q0957 gravitationally lensed quasar for
10 consecutive nights in January 2000. The resulting A image brightness curve
has significant brightness fluctuations and makes a photometric prediction for
the B image light curve for a second campaign planned for 12-21 March 2001.
The ultimate purpose is to determine the gravitational lens time delay to a
fraction of an hour, and to seek evidence for rapid microlensing.

\end{abstract}
\section{Introduction}\label{sec:Intr}

The Q0957 system is the first identified multiply imaged quasar (Walsh,
Carswell, and Weymann, 1979,) and the first to have a measured time delay
(Schild and Cholfin, 1986). It is also the first in which a microlensing event
was detected (Grieger, Kayser \& Refsdal 1988). Subsequent decades have
produced refinement of the models that describe the lensing and the many
observational parameters required to turn the physical configuration into an
important cosmological tool.

Along this path, the biggest surprise was the observation that the system
seems to show a rapid microlensing of low amplitude (Schild and Smith 1991;
Schild 1996). If confirmed, this would have important implications for the
nature of the dark matter, since solar mass objects in lens galaxy G1
should have a microlensing time scale of 30 years. It would also have
implications for the existence of fine structure in the luminous structure
of the source quasar.

Unfortunately the observational tests for this reported fine structure require
high photometric precision and an accurate value of the time delay. Worse, an
investigation by Colley and Schild (2000) in which the quasar was intensively
observed throughout 5 nights in 1994.9 and again in 1996.1 showed that from one
observatory one cannot easily measure the time delay to better than a day,
because all but the most sophisticated statistics would favor a delay where
there is no data overlap between the A and delayed B images, namely when the
time delay in an integer number of days plus a half day (the sun is up in
either the A or later B data).  Furthermore, because of this, it is virtually
impossible to assess whether or not there is microlensing on a timescale of a
day or less.  But there is good news; the quasar was demonstrated to show
brightness fluctuations of order 1 percent in a night.

Therefore, to serve two purposes, a precise determination of the time delay and
a detection of rapid microlensing, an observing campaign enlisting 10
observatories undertook continuous monitoring of the quasar images on the
nights of 20 - 29 Jan, 2000.  The present manuscript describes the observations
and results of the first year of observing, the image A component of which
becomes the prediction for the image B light curve for the coming session in
12-21 March, 2001, if there is no microlensing.

\section{Observations}

To create a continuous brightness record for Q0957+561A,B, observations from
northern hemisphere observatories would be required during their winter months.
Thus, poor weather would be likely at any site, and some redundancy would be
necessary to avoid significant gaps in the data record.  We tried to get at
least 2 participating observatories in differing weather zones for any UT hour
of the program.  Each observatory is presumed to provide about 8 hours of
coverage, which allows for substantial overlap in coverage from Eastern Asia to
Western Europe, and even into North America, though the Pacific is singly
covered by Mt. Hopkins (Arizona) and South Korea.

Our list of participating observatories is given in Table~\ref{obstab}, where
successive columns give the observatory name, telescope aperture, altitude,
latitude, and longitude, and the pixel scale of the CCD camera. Observers were
instructed to begin and end the observational session with an R filter, and to
do about 15\% of the observing through the night at V. The V data have not yet
been reduced.

Data were reduced following standard procedures and the precepts of Colley and
Schild (2000). With data from many observatories available, a standardized
header file and byte order had to be established, and the data were bias
subtracted and flat fielded with IDL software.  Simple aperture photometry,
with aperture diameters of six arcseconds, was then carried out on all image
frames.

Because the plate scales and CCD sizes of the several observatories varied
significantly, it was not possible to use exactly the same standards for all
data frames.  The data frames from Maidanak, for instance, contain only one of
the usual standards (star 5 from Schild \& Cholfin 1986).  Therefore, each of
the observatories' datasets is calibrated independently, by the Honeycutt
(1992) method for ensemble photometry.

During the aperture photometry, the FWHM seeing is calculated by using a
best-fit Moffat ($\alpha = 2.5$) profile, which allows for ellipticity and
rotation.  Running hourly averages are computed for the image A and B
photometry from each observatory.  Subtracting this average from the individual
photometric measurements from each image gives a deviation as a function of
seeing.  Seeing is expected to introduce correlated errors in the photometry,
slightly more in A than in B (e.g. Colley \& Schild 1998).  We find exactly
that behavior here, but to a differing degree from each observatory.  A
parabola accurately reproduces the behavior of the deviation from the hourly
average vs. the log of FWHM seeing.  A ``fixed'' version of the photometry can
be created by subtracting this behavior out for each photometric datum.

\section{Data Combiniation and Results}

\begin{figure}[t]
\plotone{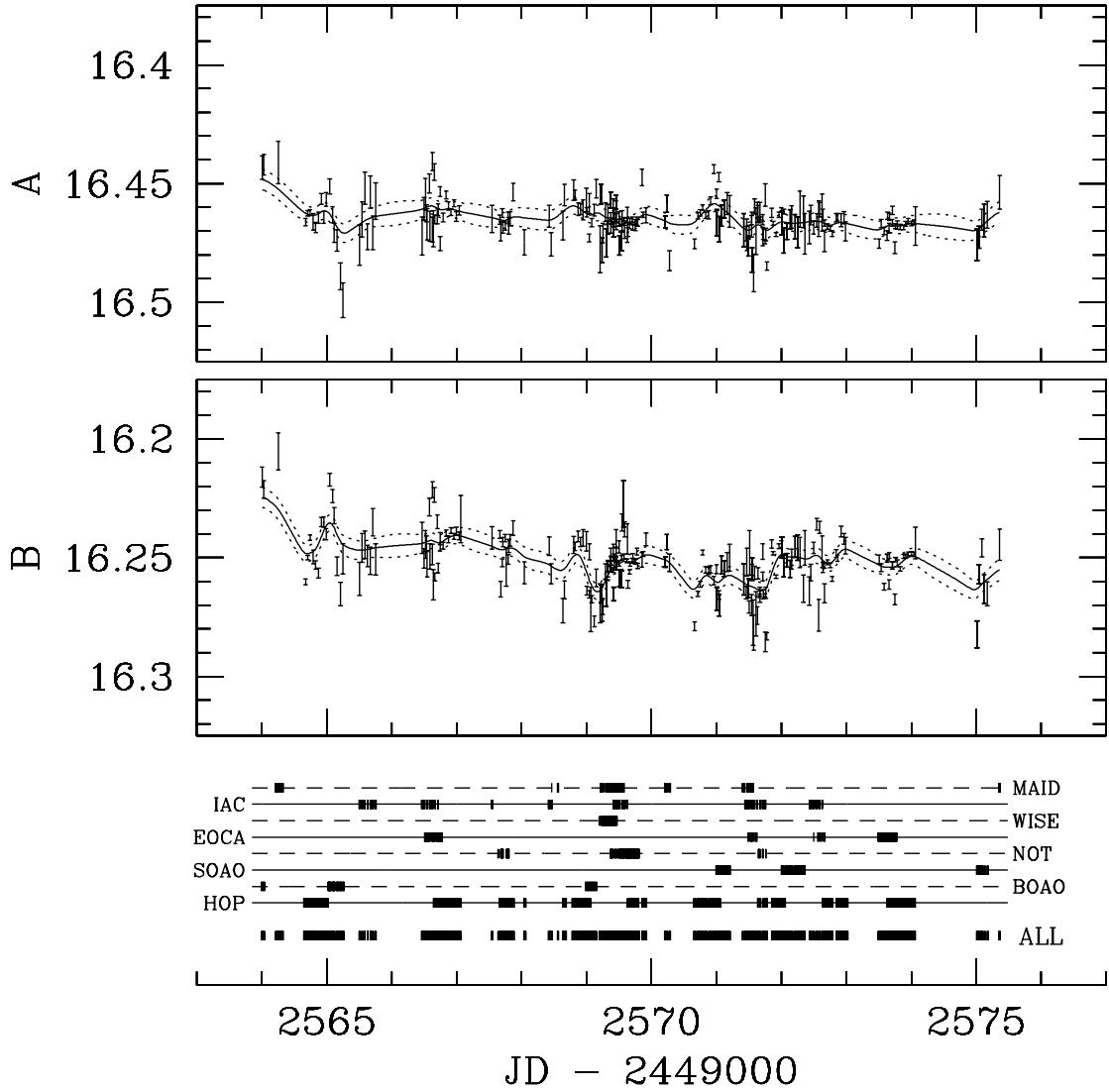}
\caption{The R-band light curve of Q0957+561A,B from January 20 to January 31,
2000.  At top is the image A brightness record; in the middle is the image B
brightness record, and at bottom is a series of line density graphs
illustrating when each observatory was contributing data.}
\label{fig1}
\end{figure}

Our R filter brightness curve is shown in Fig.~\ref{fig1}, with hourly averages
of data from each observatory.  Overplotted is the ``snake'' from the Press,
Rybicki, and Hewitt (1992) method to average and interpolate time-series
observations.  We explicitly use the structure function determined in Colley
and Schild (2000) for the interpolation.  The resultant curve is useful for our
talking and planning purposes, but the original hourly averages with errorbars
will be used in the final cross-correlation when the second year's data are
available. 
 
In combining data from several observatories we had to make adjustments, or
zero-point shifts, for each observatory.  Although the quasar brightness is
always measured relative to the standard stars, the standards are all redder
than the quasar and a small correction to the standard Kron-Cousins R filter
system is required. Such a correction is required for the Mt. Hopkins system as
well, and in practice we have determined the correction relative to the Mt
Hopkins filter/CCD response.
 
We have determined the zero-point correction by two methods.  In the first, the
measured standard star brightness is averaged for all data frames from a
particular observatory, and the average observed brightnesses compared to the
standard values plotted as a function of color of the standards. This gives a
standard curve that can be extrapolated to the colors of the quasar images to
derive the correction to the observatory's zero point. This is a normal
procedure in transforming photometry to a standard system, and we have employed
what is often referred to as first order transformation coefficients.  The
amplitude of the correction is ordinarily a few hundredths of a magnitude. For
example, the correction for the Mt Hopkins data is 0.04 mag.
 
In a second procedure, a running hourly average of the ``fixed'' data was
computed for each observatory.  We then generated the Press, Rybicki \& Hewitt
(1992) ``snake'' for all observatories except one, and computed the optimum (in
the weighted least-squares sense) offset between the snake and the photometry
from the remaining observatory.  Rotating through all the various
observatories, offsetting, then iterating, quickly produces optimum offsets
that minimize the differences in the photometry measured from all of the
observatories. 

The first and second methods produce similar results, but the first, of course,
generates an offset to be applied simultaneously to the A and B image data.
The offsets, however, would not necessarily be identical, because the lens
galaxy contributes a very red component to the very blue QSO images in very
different amounts, A to B.  This is troublesome, because one likely culprit for
non-zero photometric offsets in the first place is differences in filters and
detector responses.

So, in our second, more empirical method, we allowed the image A and image B
offsets to be determined independently.  The good news is that while the
offsets ranged over a tenth of a magnitude, they generally agreed between A and
B to better than 1\%, for all except Wise and BOAO data.  Furthermore, the
offsets from the transformation coefficients agreed with the second, empirical
method with an r.m.s. of 6 millimagnitudes (again excluding Wise and BOAO).
The Wise and BOAO offsets agree more poorly because both of those observatories
contributed principally in one night, and in both cases, showed substantial
variation (of order a few percent).  These larger variations confuse the offset
optimization (in a way that is complementary to the problem that a lack of
variations to confuse time delay estimates).  We have chosen simply to stay
with the second method (snake residual minimization), for the simple reason
that it is the more straight-forward of the two, and we invite those interested
in such nuances to consult our full data table.

We have then the problem that we have data of non-uniform quality, in
non-uniform pass-bands, zero-pointed by non-uniform standards, none of which
has a spectrum similar to our object, and none of which is even as blue as our
object.  We found, however, that by examining the colors of the standards, we
could generate approximate offsets that agree quite well with completely
empirical offsets, and that the offsets themselves are quite consistent between
the A and B images at a single observatory.  We therefore adopt this empirical
method as our best procedure for determining offsets.
 
We are pleased that our image A brightness curve shows some brightness
fluctuations that can be sought in our forthcoming second season of observing.
Relative to simple weighted mean for the entire data record, we found that the
$\chi^2$ value is greater than 20,000 for 296 degrees of freedom for both A and
B images.  The non-uniformity of the data makes such a direct calculation
assumptive, so we also computed this $\chi^2$ value for the Mt. Hopkins data,
alone, in which we found values of greater than 1000 for 93 degrees of freedom.
While this is certainly not a detailed statistical treatment, our previous work
(Colley \& Schild 1999) showed that our errorbars on Mt. Hopkins data were
quite reliable, and we therefore have confidence that this measurement shows
siginificant departures from constant.

Furthermore, the complex pattern of brightness fluctuations seen around
$\mbox{JD}-2449000$ = 2569 -- 2572 is fairly well defined by observations and
an even larger, though more poorly constrained, drop is evident at 2565.3.
This latter feature is well sampled from observations in Korea, but we remain
cautious, because the drop occurs simultaneously in both images, which seems
unlikely.  We have checked the data with another software program (IRAF) and
find no fault with the reductions.  If the feature is real, it gives an
important feature to be sought near the beginning of the observational period
now being planned. If this feature can be identified on 12 March 2001,
determination of a time delay to less than an hour should be possible.

Data tables, both before and after binning, may be found at URL:\\
http://cfa-www.harvard.edu/$\sim$wcolley/Q0957/data/.

\begin{acknowledgements}
We thank Prof. Nail Sakhibullin and Dr. Ilfan Bikmaev of Kazan State University
for telescope time allocation and support.  We thank the Maidanak Foundation
for purchasing the ST-7 camera for Maidanak Observatory.  We are also grateful
to the administration of the Ulugh Beg Astronomical Institute of Ac. Sci. and
to its director, Prof. Ehgamberdiev.  Nordic Optical Telescope is operated on
the island of La Palma jointly by Denmark, Finland, Iceland, Norway and Sweden,
in the Spanish Observatorio del Roque de los Muchachos of the Instituto de
Astrof\'{\i}sica de Canarias.
\end{acknowledgements}

\begin{table}
\begin{tabular}{cccccc}
\tableline
Observatory & Aperture & Altitude & Latitude & Longitude & CCD \\
            & (meters) & (meters) &          & (hours)   & $N_x \times N_y$,
arcsec/pix \\
\tableline
\tableline

Mt. Hopkins, USA & 1.2 & 2608 & $31\deg 41\arcmin$ & -7.4 & $2048 \times 2048$,
0.31 \\
NOT, Canary Is.         & 2.5 & 2395 & $28\deg 17\arcmin$ & -1.1 & $2048 \times
2048$, 0.2 \\
JKT, Canary Is.         & 1.0 & 2395 & $28\deg 17\arcmin$ & -1.1 & $2048 \times
2048$, 0.33 \\
IAC 80, Canary Is.      & 0.8 & 2400 & $28\deg 18\arcmin$ & -1.1 & $1024 \times
1024$, 0.43 \\
EOCA, Spain        & 1.52& 2168 & $37\deg 13\arcmin$ & -0.2 & $1024 \times
1024$, 0.4 \\ 
TUG, Turkey     & 1.5 & 2550 & $36\deg 50\arcmin$ & +2.0 & $1530 \times
1020$, 0.16 \\
Wise, Israel        & 1.0 & 874  & $30\deg 36\arcmin$ & +2.3 & $500 \times
500$, 0.7 \\
Maidanak, Uzbekistan   & 1.5 & 2600 & $38\deg 41\arcmin$ & +4.5 & $765 \times
510$, 0.15 \\
BOAO, S. Korea        & 1.8 & 1162 & $36\deg 10\arcmin$ & +8.5 & $2048 \times
2048$, 0.34 \\
SOAO, S. Korea        & 0.61& 1378 & $36\deg 56\arcmin$ & +8.6 & $2048 \times
2048$, 0.5 \\

\tableline
\end{tabular}
\caption{Observatories in the first season of the ``QuOC Around the Clock''
Collaboration.}
\label{obstab}
\end{table}

\end{document}